\documentclass[a4paper]{article}
\usepackage{amsmath}
\usepackage{graphicx}
\usepackage{geometry}
\usepackage{floatrow}
\usepackage{layout}
\usepackage{amssymb} 
\usepackage{multirow}
\usepackage{caption}
\usepackage{authblk}
\usepackage{indentfirst}
\usepackage[hidelinks]{hyperref}
\usepackage{longtable}
\usepackage{caption}
\usepackage{subcaption}
\usepackage{appendix}
\usepackage{algorithm}
\usepackage{algpseudocode}
\usepackage{subcaption}

\providecommand{\keywords}[1]
{
  \small	
  \textbf{\; \textit{Keywords---}} #1
}

\begin{document}

\title{\textbf{\huge{Building a Recommendation System Using Amazon Product Co-Purchasing Network}}}

\author{\textbf\large{Mark Liu, Catherine Zhao, Nathan Zhou}}

\affil{\textbf{Washington University in St. Louis}}

\date{\today}

\maketitle
\begin{abstract}
This project aims to develop an online, inductive recommendation system for newly listed products on e-commerce platforms, focusing on suggesting relevant new items to customers while they are purchasing other products. Using the "Amazon Product Co-Purchasing Network Metadata" dataset \cite{10.1145/1232722.1232727}, we construct a co-purchasing graph where nodes represent products and edges capture co-purchasing relationships between them. To address the challenge of recommending new products with limited information, we apply a modified GraphSAGE method for link prediction. This inductive approach leverages both product features and the existing co-purchasing graph structure to predict potential co-purchasing relationships, allowing the model to generalize to unseen products. As an online method, it can update in real-time, making it scalable and adaptive to evolving product catalogs. Our results demonstrate that this approach outperforms baseline algorithms in predicting relevant product links, offering a promising solution for enhancing the relevance of new product recommendations in e-commerce environments. All the codes are available at \href{https://github.com/cse416a-fl24/final-project-l-minghao_z-catherine_z-nathan.git}{\texttt{https://github.com/cse416a-fl24/final-project-l-minghao\_z-catherine\_z-nathan.git}}.


Resources:
Data: The "Amazon Product Co-Purchasing Network Metadata" \cite{10.1145/1232722.1232727}, comprising product metadata for 548,552 products across various categories (e.g., books, CDs, DVDs).
Software: Python, NetworkX for graph analysis, Scikit-learn and PyTorch for machine learning, and ScentenceTransformer for title embeddings.
\end{abstract}


\section{Introduction}
E-commerce platforms like Amazon use recommendation systems to help customers discover new products, boosting sales and enhancing user experience. Traditional methods, such as collaborative filtering, often struggle with the New Item Problem, where new products lack sufficient data for accurate recommendations. These systems also face challenges with scalability as product catalogs grow.

To overcome these limitations, we propose leveraging the co-purchase network of products, where relationships between items are captured in a graph. By using the GraphSAGE method \cite{hamilton2018}, which allows for scalable, inductive learning from large graphs, we can predict product links even when limited information is available. This approach combines product features with network structure to improve recommendations for new items based on their co-purchasing relationships with existing popular products.

The goal of this project is to predict links for isolated nodes (new products) in the Amazon Product Co-Purchasing Network dataset. We hypothesize that:

\begin{itemize}
    \item \textbf{H1:} Recommending new products based on predicted co-purchasing links with popular items will make the recommendations more relevant.
    \item \textbf{H2:} Using 1-degree neighbors in the co-purchase network will provide accurate ground truth for link prediction.
    \item \textbf{H3:} The graph structure around targeted products will improve link prediction accuracy.
\end{itemize}

By integrating these insights, we aim to build a scalable recommendation system that enhances the relevance of new product suggestions.

\section{Data}
\subsection{Network Creation}
\begin{longtable}{|c|c|}
\caption{Dataset Features (Features highlighted in red will be used.)} \label{tab:dataset_features} \\  
\hline
\textbf{Feature} & \textbf{Description} \\ \hline
\endfirsthead
\hline
\textbf{Feature} & \textbf{Description} \\ \hline
\endhead
\hline
\endfoot
\hline
\endlastfoot

\textbf{Id} & Product ID (numbers from 0 to 548551) \\ \hline
\textcolor{red}{\textbf{ASIN}} & Amazon Standard Identification Number \\ \hline
\textcolor{red}{\textbf{Title}} & Name or title of the product \\ \hline
\textcolor{red}{\textbf{Group}} & Product group (e.g., Book, DVD, Video, Music) \\ \hline
\textbf{Salesrank} & Amazon Salesrank, indicating product popularity \\ \hline
\textcolor{red}{\textbf{Similar}} & ASINs of co-purchased products (e.g., products often bought together) \\ \hline
\textcolor{red}{\textbf{Categories}} & Product categories in Amazon’s hierarchy (e.g., Books|Fiction|Mystery) \\ \hline
\textbf{Reviews} & Product review data (rating, total number of votes, and helpfulness votes) \\ \hline

\end{longtable}

The dataset used is the "Amazon Product Co-Purchasing Network Metadata" \cite{10.1145/1232722.1232727} from SNAP, which contains metadata and review information for 548,552 products, as shown in Table \ref{tab:dataset_features}. We created a network where each product is a node, with ASIN, Title, Group, and Categories as node features. Edges represent co-purchase relationships based on the "Similar" feature. The dataset was preprocessed to remove products with missing information. 

\subsection{Key Network Statistics and Visualizations}

The network after preprocessing has  964,468 edges and 519,497 nodes, of which 159,575 are isolated. We worked on a undirected graph, as product A being co-purchased with B implies B being co-purchased with A. The first statistic we calculate is the number of connected components. We get that there are 165,510  connected components, of which 5,935 are not isolated nodes.

\begin{figure}[!htbp]
    \centering
    \includegraphics[width=0.6\linewidth]{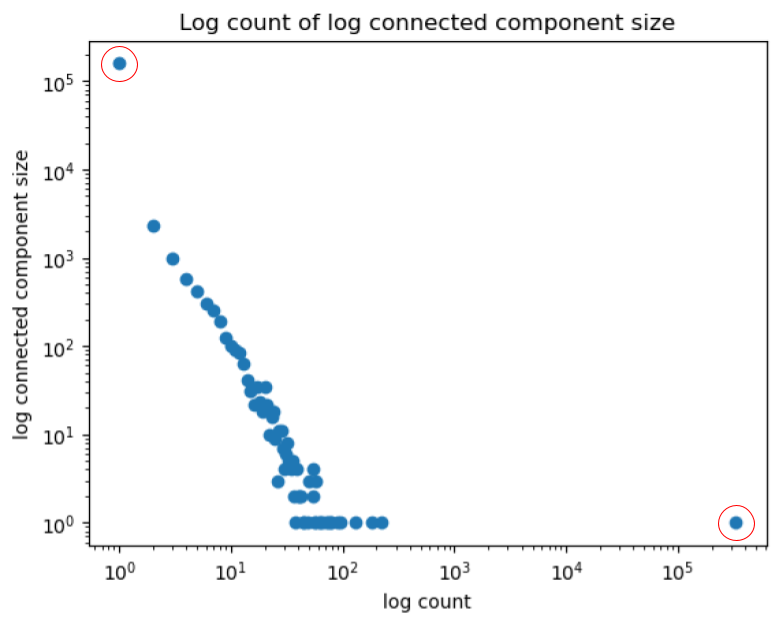}
    \caption{Plot of log connected component size with log frequency}
    \label{fig:enter-label}
\end{figure}
To reduce training time and avoid issues with isolated nodes lacking network structure and ground truth for link prediction, we extracted the largest connected component (LCC) from the dataset. This subgraph contains 327,953 nodes and 902,604 edges. The associativity coefficient for groups within the LCC is 0.327, suggesting that products tend to be co-purchased with others of the same type, such as books with other books. When plotting the degree distribution of the entire graph and the LCC on a log-log scale, both exhibit a power law distribution. Since the LCC has no nodes with degree 0, we used the complementary cumulative distribution function (CCDF) to estimate the coefficient of the power law distribution, which is found to be 3.55.

\begin{figure}[!htbp]
\centering
\begin{subfigure}{.5\textwidth}
  \centering
  \includegraphics[width=.8\linewidth]{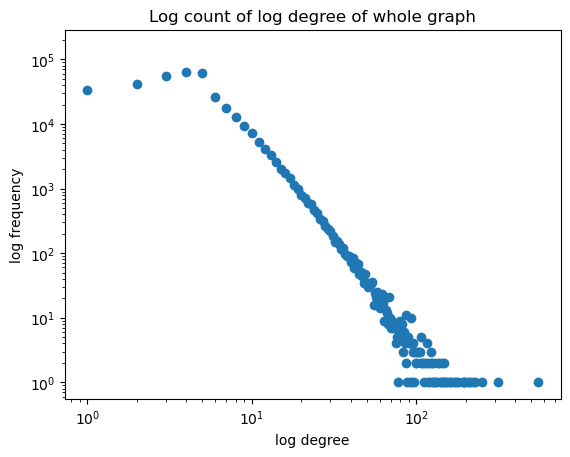}
  \caption{Log-Log distribution of entire graph}
  \label{fig:sub1}
\end{subfigure}%
\begin{subfigure}{.5\textwidth}
  \centering
  \includegraphics[width=.8\linewidth]{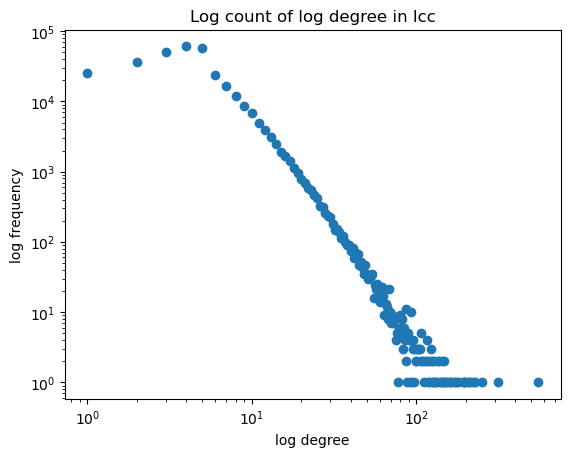}
  \caption{Log-Log distribution of LCC}
  \label{fig:sub2}
\end{subfigure}
\caption{Degree Distirbution Scatter Plots}
\label{fig:test}
\end{figure}
We applied the Louvain Community Detection Algorithm to the LCC to gain deeper insights into its structure. This algorithm was chosen for its efficiency in large networks due to its relatively low time complexity. The modularity score of the partition was found to be 0.926, indicating strong community structure. This is further validated by the presence of bridges when visualizing the network. To create a clear visualization, we selected the top 50 highest-degree nodes, all located within the LCC, along with their neighbors, and imported the subgraph into Gephi. By zooming in on a specific section of the graph as showed in \ref{fig:enter-label}, we observed a local bridge connecting two distinct communities.

\begin{figure}[!htbp]
    \centering
    \includegraphics[width=0.6\linewidth]{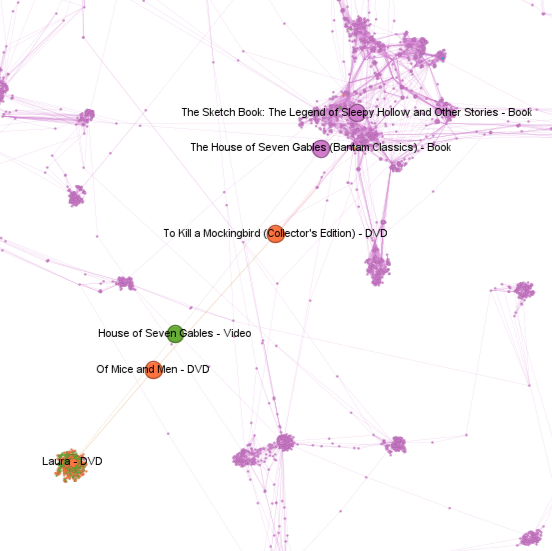}
    \caption{Zoomed in visualization of top 50 nodes and their neighbors. Node color represents the group the node belongs to. Node size of bridge has been increased for visibility. }
    \label{fig:enter-label}
\end{figure}

\section{Methodology}
\subsection{Network Analysis Approach}
We will use \textbf{Degree Centrality} and \textbf{Clustering Coefficient} as centrality measures. The \textbf{Louvain Modularity method} is used to detect communities by iteratively merging nodes to maximize modularity. Note that the \textbf{Modularity Score} based on a single attribute is given by:
\[
Q = \frac{1}{2m} \sum_{i,j} \left[ A_{ij} - \frac{k_i k_j}{2m} \right] \delta(c_i, c_j)
\]

\subsection{Node Embeddings}
The features for each product are encoded as follows: \textbf{Title Embeddings} \( T(u) \) are generated by encoding the product titles using a pre-trained language model (e.g., SentenceTransformer \cite{reimers-2019-sentence-bert}) and reducing the dimensionality to 32 using PCA. \textbf{Group Embeddings} \( G(u) \) are created using one-hot encoding to convert product groups (e.g., "Book", "DVD", "Video", "Music") into binary vectors. The \textbf{Category Vector} \( CV(u) \) is constructed by extracting and parsing category IDs from each node's hierarchical category string using regular expressions. Additionally, \textbf{Node Degree} \( D(v) \) and \textbf{Clustering Coefficient} \( CC(v) \) are computed for each node. We also calculate the \textbf{categorical similarity} \( Sim_c(u,v) \) between a pair of nodes, which is represented as a binary vector that highlights the deepest common category in their padded category sequences, with matches marked as 1 and mismatches as 0. For further details, see Algorithm \ref{alg:extract_category_similarity}.

\subsection{Training Dataset Creation} As hypothesized in H2, 1-degree nodes and their neighbors serve as the ground truth for link prediction in isolated nodes. The training dataset is then created by selecting positive and negative samples. For positive samples, we choose 10,000 1-degree nodes along with their neighbors, labeling them as 1. For negative samples, we select 10,000 1-degree nodes and pair them with nodes that have no connections, labeling them as 0.

\subsection{Baseline Model}
A \textbf{Random Forest Classifier} \cite{breiman2001random}, which is an ensemble learning method that constructs 100 decision trees and aggregates their predictions to enhance accuracy, will be used as the baseline model. The input features for a training pair $(u, v)$ are represented as $[T(u), G(u), T(v), G(v), Sim_v(u,v), D(v), CC(v)]$, where the structural features of the target node $v$ are implicitly incorporated.

\subsection{GraphSAGE Model Overview and Modifications}

\subsubsection{GraphSAGE Model Overview}

\textbf{GraphSAGE \cite{hamilton2018}} (Graph Sample and Aggregation) is a graph neural network (GNN) algorithm that generates node embeddings by sampling and aggregating features from a node's neighbors. Unlike traditional methods that require the entire graph for training, GraphSAGE uses a sampling-based approach, making it scalable to large graphs. It learns an aggregation function (e.g., mean, max) to combine features from neighbors, enabling the model to generalize to unseen nodes. This makes GraphSAGE particularly suitable for real-time applications, such as large-scale recommendation systems.

\subsubsection{Modified GraphSAGE for Link Prediction in Co-Purchasing Network}

In our case, we adapt \textbf{GraphSAGE} to address specific challenges that arise when applying it to a co-purchasing network, especially for \textit{isolated nodes} (nodes with no direct neighbors). While GraphSAGE traditionally aggregates information from a node's neighbors to create embeddings, isolated nodes do not have any neighbors and thus lack the local structure typically used for learning. To overcome this limitation, we propose the following modifications:

\begin{itemize}
    \item \textbf{Node Pair Embedding as Feature}: Instead of relying solely on neighbor aggregation, we treat the feature between the pair of nodes as node embeddings. For pairs of nodes, we sample from the target node's neighbors and use these as a proxy for aggregation.
    \item \textbf{One-Hop Sampling}: In a co-purchasing network, \textit{two-hop sampling} (e.g., using indirect neighbors) is not suitable. For example, if product \(i_1\) is co-purchased with \(i_2\) and \(i_2\) is co-purchased with \(i_3\), this does not imply that \(i_1\) and \(i_3\) will co-purchase. Therefore, we limit our sampling to \textit{one-hop neighbors} only, ensuring that only directly connected products are used to predict future co-purchases.
\end{itemize}

By introducing these modifications, our approach ensures that \textbf{GraphSAGE} can still be applied to isolated nodes in the co-purchasing network, leveraging the local neighborhood information and graph structure to make relevant predictions even in the absence of direct neighbors. This modified version aligns with our hypothesis (H3), which suggests that \textbf{graph structure} can still improve link prediction for isolated nodes.

\section{Analysis, Results, and Discussion Section}
\subsection{Preliminary Study}
Modularity scores are calculated for the communities identified by the Louvain method, grouped according to the group attribute and category similarity, as shown in Table \ref{tab:modularity_scores}. While the community structure is robust for general connectivity, it is weak for specific attributes, suggesting that relying on a single attribute is not an effective indicator of community structure. As a result, grouping isolated nodes with those sharing similar attributes may not lead to optimal performance.

\begin{table}[h!]
\centering
\begin{tabular}{|c|c|}
\hline
\textbf{Modularity Score Type} & \textbf{Score} \\ \hline
Louvain Community Modularity Score & 0.926 \\ \hline
Modularity Score by Category Similarity & 0.155 \\ \hline
Modularity Score by Group Attribute & 0.181 \\ \hline
\end{tabular}
\caption{Modularity Scores for Community Detection}
\label{tab:modularity_scores}
\begin{flushleft}
\textit{Note:} Modularity Score for category similarity is modified by substituting \( \delta(c_i, c_j) \) with \( \frac{\sum \text{Sim}_c(i,j)}{\text{len}(\text{Sim}_c(i,j))} \).
\end{flushleft}
\end{table}

\subsection{Ablation Experiment}
An ablation experiment was conducted using the Random Forest Classifier to assess the contribution of each attribute to the final result. As shown in Table \ref{tab:ablation_study}, the Title Embedding had minimal impact on performance. This is reasonable, as titles for products like DVDs and books can barely capture the content information. Therefore, we decided to remove it to improve efficiency.

\begin{table}[htbp]
\centering
\caption{Ablation Study Results}
\label{tab:ablation_study}
\begin{tabular}{|c|c|c|c|c|c|}
\hline
\textbf{Feature Variant} & \textbf{Precision} & \textbf{Recall} & \textbf{F1 Score} & \textbf{ROC AUC} \\ \hline
\textbf{Baseline} & 0.9137 & 0.9128 & 0.9133 & 0.9709 \\ \hline
\textbf{Without Title Embeddings} & 0.9069 & 0.9118 & 0.9094 & 0.9667 \\ \hline
\textbf{Without Group Vectors}& 0.8919 & 0.8826 & 0.8872 & 0.9607 \\ \hline
\textbf{Without Category Similarity} & 0.8416 & 0.9237 & 0.8808 & 0.9422 \\ \hline
\textbf{Without Degree Features} & 0.8982 & 0.8737 & 0.8858 & 0.9541 \\ \hline
\textbf{Without Cluster Features} & 0.8958 & 0.8816 & 0.8887 & 0.9606 \\ \hline
\end{tabular}
\end{table}

\subsection{Results}

To compare the models' performance, we test them on a subgraph sampled as follows:  
We randomly select a start node from the largest connected component, then perform a breadth-first search (BFS) to explore and collect connected nodes until the desired number of samples (e.g., 1000) is reached. This method ensures that the sampled nodes are connected, preserving the local graph structure, and helps maintain the relationships between the nodes within the subgraph. We use top-k accuracy as the metric, where a prediction is considered correct if the ground truth is among the top-k nodes with the highest probability. Since our model outperforms the other models in almost all $k$' s, and especially in the 300-400 range, this means that our model is much more likely to have good recommendations suggested with a higher probability, thus decreasing the manual verification time of the recommendations if used in the real world. Moreover, this means that reasonably high certainty can be given to much smaller potential reconmendation lists. The result is plotted in Figure \ref{fig:top-k_accuracy}.

\subsection{Discussion}
The results show that the modified GraphSAGE outperforms the baseline model in predicting co-purchasing links for isolated nodes. Its capacity to generalize to unseen nodes without the need for retraining demonstrates its practical value for dynamic e-commerce platforms. While the Modified GraphSAGE is still based on a binary classifier, it surpasses the Random Forest Classifier in modeling probabilities.
\section{Conclusion}
This project successfully developed a recommendation system leveraging co-purchasing relationships to suggest new products. Our modified GraphSAGE model, which explicitly utilizes graph structure, outperforms the baseline model while preserving scalability and inductive capabilities, making it well-suited for real-time recommendations.\\

Future work will focus on incorporating customer feedback, such as average product ratings and sales rank. By integrating these metrics, we aim to create a hierarchical ranking framework for selecting recommendation candidates.
\newpage
\bibliography{references}

\begin{thebibliography}{1}

\bibitem{10.1145/1232722.1232727}
J.~Leskovec, L.~A. Adamic, and B.~A. Huberman, ``The dynamics of viral marketing,'' {\em ACM Trans. Web}, vol.~1, p.~5–es, May 2007.

\bibitem{hamilton2018}
W.~L. Hamilton, R.~Ying, and J.~Leskovec, ``Inductive representation learning on large graphs,'' 2018.

\bibitem{reimers-2019-sentence-bert}
N.~Reimers and I.~Gurevych, ``Sentence-bert: Sentence embeddings using siamese bert-networks,'' in {\em Proceedings of the 2019 Conference on Empirical Methods in Natural Language Processing}, Association for Computational Linguistics, 11 2019.

\bibitem{breiman2001random}
L.~Breiman, ``Random forests,'' {\em Machine Learning}, vol.~45, no.~1, pp.~5--32, 2001.

\end{thebibliography}
\bibliographystyle{ieeetr}

\begin{appendices}
\section{Extract Category Similarity Function}

Let \( C_1 \) and \( C_2 \) be the category lists for nodes \( \text{node}_1 \) and \( \text{node}_2 \), respectively. The goal is to compute the category similarity for each pair of nodes \( (\text{node}_1, \text{node}_2) \).

\begin{algorithm}[htbp]
\caption{Extract Category Similarity}\label{alg:extract_category_similarity}
\begin{algorithmic}[1]
\State \textbf{Input:} $samples$, $category\_index$
\State \textbf{Output:} $category\_similarity$
\State Initialize $category\_similarity \gets []$
\For{each $(node1, node2)$ in $samples$}
    \State $max\_index \gets -\infty$
    \For{each $c1$ in $category\_index[node1]$}
        \For{each $c2$ in $category\_index[node2]$}
            \State $common\_categories \gets \text{compare}(c1, c2)$
            \If{any common categories}
                \State $index \gets \text{argmax}(common\_categories)$
                \If{$index > max\_index$}
                    \State $max\_index \gets index$
                \EndIf
            \EndIf
        \EndFor
    \EndFor
    \State Append similarity to $category\_similarity$
\EndFor
\State \Return $category\_similarity$
\end{algorithmic}
\end{algorithm}

\section{Results}
\begin{figure}[htbp]
    \centering
    \includegraphics[width=0.8\linewidth]{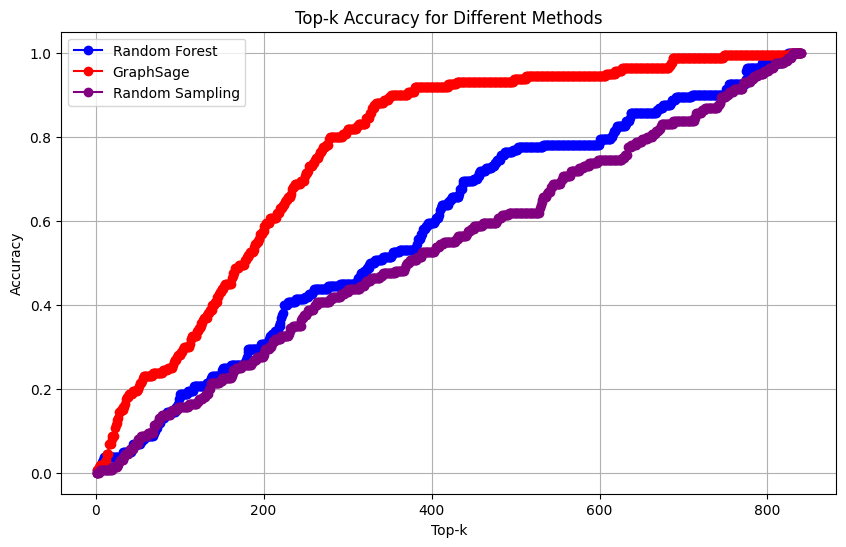}
    \caption{Top-k accuracy for Modified-GraphSAGE, Random Forest, and Random Choice. Top-5 Accuracy: Random Forest (0.0125), GraphSAGE Top-5 (0.0187), Random Sampling (0.0063).}
    \label{fig:top-k_accuracy}
\end{figure}

\begin{figure}[htbp]
    \centering
    \begin{minipage}{0.49\linewidth}
        \centering
        \includegraphics[width=\linewidth]{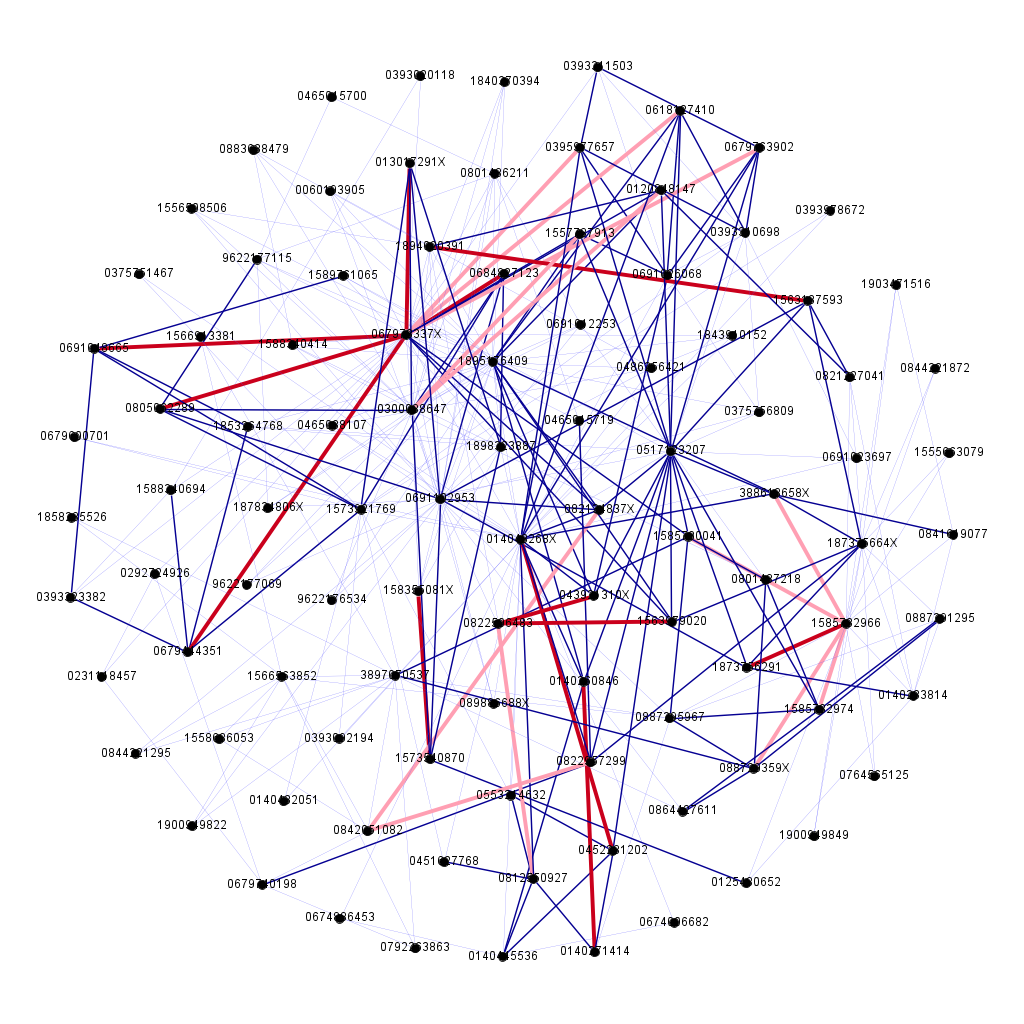}
        \subcaption{Modified-GraphSAGE Predictions}
        \label{fig:prediction_GS}
    \end{minipage}
    \hfill
    \begin{minipage}{0.49\linewidth}
        \centering
        \includegraphics[width=\linewidth]{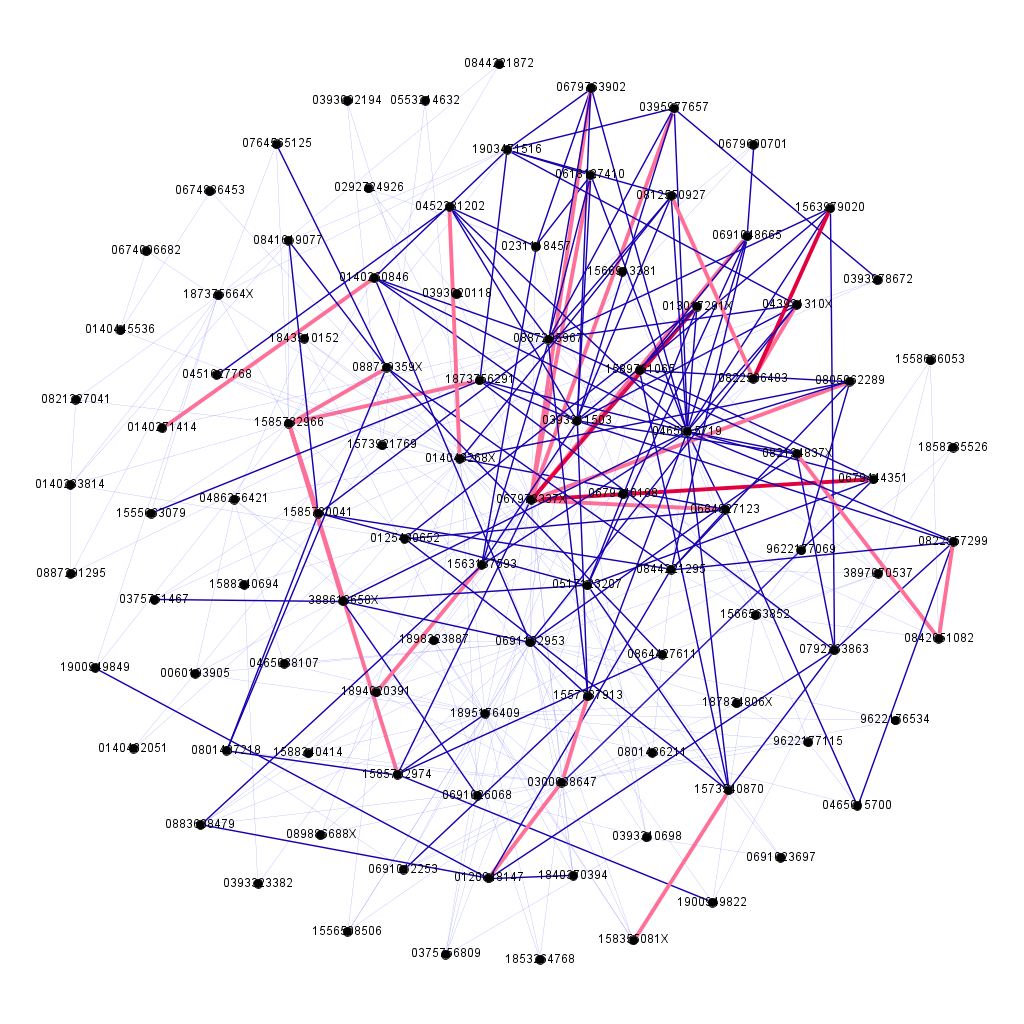}
        \subcaption{Random Forest Predictions}
        \label{fig:prediction_RF}
    \end{minipage}
    \caption{A sampled visualization for predictions in a subgraph with 100 nodes: Dark blue edges are top-3 candidate nodes selected by the isolated nodes, pink edges point to ground truth targets, and the red one represents a successful prediction.}
    \label{fig:predictions_comparison}
\end{figure}

\end{appendices}

\end{document}